\documentclass[12pt]{article}
\usepackage{makeidx}
\usepackage{amssymb}
\usepackage{amsfonts}
\usepackage{amsmath}
\usepackage{graphicx}
\usepackage{setspace}
\usepackage{authblk}
\usepackage{units}
\usepackage{appendix}
\newcommand{\lambdabar}{{\mkern0.75mu\mathchar '26\mkern -9.75mu\lambda}}
\usepackage[left=2cm,top=2cm,right=2cm,bottom=2cm]{geometry}

\begin{document}

\title{\textbf{Coherent and semiclassical states of a charged particle in a constant
electric field}}
\author[1]{T. C. Adorno\thanks{tg.adorno@gmail.com, tg.adorno@mail.tsu.ru}}
\author[2]{A. S. Pereira\thanks{albertoufcg@hotmail.com, spa3@tpu.ru.}}
\affil[1]{\textit{Department of Physics, Tomsk State University, Lenin Prospekt 36, 634050, Tomsk, Russia;}}
\affil[2]{\textit{Department of Higher Mathematics and Mathematical Physics,
\ Institute of Physics and Technology, Tomsk Polytechnic University, Russia}}

\maketitle

\onehalfspacing

\begin{abstract}
The method of integrals of motion is used to construct families of generalized coherent states of a nonrelativistic spinless charged particle in a constant electric field. Families of states, differing in the values of their standard deviations at the initial time, are obtained. Depending on the initial values of the standard deviations, and also on the electric field, it turns out to be possible to identify some families with semiclassical states.


\end{abstract}

\section{Introduction\label{Sec1}}

Quantum states minimizing the uncertainty relation for a pair of observables, for example, the position and momentum, were proposed in 1926 by Schr{\"o}dinger
\cite{Sch1926}. Since the appearance of lasers in the 1960s, these states have
attracted attention after the appearance of papers by Rashevskii \cite{Rashveskii}, Glauber \cite{Gla1963-130}, Klauder  \cite{Kla1960,Kla1963-4,Kla1968} and Sudarshan \cite{Sud1963}. The theoretical formulation lying at the basis of these works has enabled researchers to open up a new branch of quantum physics, known as quantum optics, and the term {\textit{coherent states}} (CSs) was first used in this context. For a harmonic oscillator, CSs have three equivalent definitions, namely: 1) CSs are eigenfunctions of the annihilation operator; 2) they minimize Heisenberg’s uncertainty relation; and 3) they are obtained by a shift of the vacuum state under the action of the unitary operator of the Weyl–Heisenberg group. These states form a complete, but non-orthogonal set of vectors in Hilbert space, which makes it possible to characterize them as an overcomplete basis \cite{Kla1963-4}. Because of the properties of these states, various generalizations of CSs were proposed in the semiclassical description of quantum systems, in the theory of quantization, in the physics of condensed matter, in the theory of radiation, etc. (for example, see
\cite{Kla1985,Per1986,Gaz2009,Nie2000,MalMa,DodMa87,gilm72,AAG00}).

Another important class of quantum states which enables a correspondence to a classical description is known as semiclassical states (SSs). In these states, the mean values of observables are larger than the corresponding standard deviations. The significance of this property is that the particle actually moves along classical trajectories. In particular, CSs of a harmonic oscillator coincide with SSs, provided the standard deviation does not depend on time. However, identity of CSs and SSs cannot be established in the general case. This is possible only in special cases under some conditions on the parameters. A study of these cases for a nonrelativistic spinless charged particle in a constant electric field is the subject of the present work.

In section \ref{Sec2} we provide a review of classical and quantum dynamics and briefly discuss the method of integrals of motion for the problem under consideration
\cite{DodMa87,Bag2013,Bag2014,Bag2015}. Within the framework of this method, we establish various families of CSs and {{\textit{generalized coherent states}} (GCSs), satisfying the Schr{\"o}dinger equation. For both types of states
we discuss the standard deviations and uncertainty relations. In Section \ref{Sec3}, following the model proposed in \cite{Bag2014}, we analyze GCSs and the conditions under which these states can be considered as semiclassical.

\section{Charged particle in constant and uniform electric field\label{Sec2}}

\subsection{Classical and Quantum dynamics. Coherent states\label{Sec2.1}}

Let us consider a nonrelativistic charged particle with charge $e$ (for an
electron, $e=-\left\vert e\right\vert $), interacting with a constant and uniform
electric field $\mathbf{E}=\left(  0,0,E\right)  $, with
potentials $A_{0}\left(  z\right)  $, $\mathbf{A}=\left(  0,0,A\left(
t\right)  \right)  $ given by the formulas%
\begin{equation}
A_{0}\left(  z\right)  =-zE\sin^{2}\alpha\,,\ \ A\left(  t\right)
=-ctE\cos^{2}\alpha\,,\ \ \alpha\in\left[  0,\pi/2\right]  \,. \label{1.2}%
\end{equation}
The particle moves along the $z$ axis, $z\in\left(  -\infty
,+\infty\right)  $. The Hamiltonian has the form%
\begin{align}
H  &  =\frac{1}{2m}\left(  p_{z}-\frac{e}{c}A\left(  t\right)  \right)
^{2}+eA_{0}\left(  z\right) \nonumber\\
&  =\frac{p_{z}^{2}}{2m}-m\xi z\sin^{2}\alpha+\xi p_{z}t\cos^{2}\alpha
+\frac{m\xi^{2}}{2}t^{2}\cos^{4}\alpha\,, \label{1.3}%
\end{align}
where $\xi=eE/m$. Hamilton's equations and their are%
\begin{align}
&  \dot{z}\left(  t\right)  =\frac{\partial H}{\partial p_{z}}=\frac{p_{z}}%
{m}+\xi t\cos^{2}\alpha\,,\text{ \ }\dot{p}_{z}\left(  t\right)
=-\frac{\partial H}{\partial z}=m\xi\sin^{2}\alpha\,,\nonumber\\
&  z\left(  t\right)  =z_{0}+\frac{p_{z}^{0}}{m}t+\frac{1}{2}\xi t^{2},\text{
\ }p_{z}\left(  t\right)  =p_{z}^{0}+m\xi t\sin^{2}\alpha\,, \label{1.5}%
\end{align}
where $z_{0}=z\left(  0\right)  ,\ p_{z}\left(  0\right)  =p_{z}^{0}$ are the
initial values of the Cauchy problem.

The quantum dynamics is described by the Schr{\"o}dinger equation%
\begin{align}
&  i\hbar\partial_{t}\Psi\left(  z,t\right)  =\hat{H}\Psi\left(  z,t\right)
\,,\nonumber\\
&  \hat{H}=\frac{\hat{p}_{z}^{2}}{2m}-m\xi z\sin^{2}\alpha+\xi\hat{p}_{z}%
t\cos^{2}\alpha+\frac{m\xi^{2}}{2}t^{2}\cos^{4}\alpha\,,\label{2.1}%
\end{align}
which in dimensionless quantities%
\begin{equation}
\hat{q}=l^{-1}\hat{z}\,,\ \ \hat{p}=\frac{l}{\hslash}\hat{p}_{z}%
\mathbf{\,},\ \ \left[  \hat{q},\hat{p}\right]  =i\,,\text{ \ }\tau
=\frac{\hbar}{ml^{2}}t\,,\text{ \ }\Xi=\frac{m^{2}l^{3}}{\hbar^{2}}%
\xi\,,\label{2.2}%
\end{equation}
is rewritten as%
\begin{align}
&  \hat{S}\Phi\left(  q,\tau\right)  =0,\text{ \ }\hat{S}=\partial_{\tau
}+i\hat{\mathcal{H}},\text{ \ }\Phi\left(  q,\tau\right)  =\sqrt{l}\Psi\left(
lq,\frac{ml^{2}}{\hbar}\tau\right)  \,,\nonumber\\
&  \hat{\mathcal{H}}=\frac{1}{2}\hat{p}^{2}+\Xi\tau\hat{p}\cos^{2}\alpha
-\Xi\hat{q}\sin^{2}\alpha+\frac{1}{2}\Xi^{2}\tau^{2}\cos^{4}\alpha,\text{
\ }\hat{H}=\frac{\hbar^{2}}{ml^{2}}\hat{\mathcal{H}}\,.\label{2.3}%
\end{align}

To solve the Schr{\"o}dinger equation, it is useful introduce the operator $\hat
{A}\left(  \tau\right)  $, associated with the canonical pair $\left(  \hat
{q},\hat{p}\right)  $ by the linear transformation%
\begin{equation}
\hat{A}\left(  \tau\right)  =\frac{f\left(  \tau\right)  \hat{q}+ig\left(
\tau\right)  \hat{p}}{\sqrt{2}}+\varphi\left(  \tau\right)  \,, \label{2.4}%
\end{equation}
where the coefficients $f\left(  \tau\right)  $, $g\left(  \tau\right)  $,
$\varphi\left(  \tau\right)  $ depend on time. Operator \ref{2.4} is an integral of motion \cite{DodMa87,Bag2013,Bag2014,Bag2015}
if the condition\footnote{If $\hat{H}$ is self-adjoint, the adjoint operator
$\hat{A}^{\dagger}\left(  \tau\right)  $ is an integral of motion provided
that Eq. (\ref{2.5}) is satisfied.}%
\begin{equation}
d_{\tau}\hat{A}\left(  \tau\right)  =\left[  \hat{S},\hat{A}\left(
\tau\right)  \right]  =0\,, \label{2.5}%
\end{equation}
is fulfilled. Taking condition (\ref{2.5}) into account, the
functions $f\left(  \tau\right)  $, $g\left(  \tau\right)  $, $\varphi\left(
\tau\right)  $ should satisfy the first-order differential equations%
\begin{equation}
\dot{f}\left(  \tau\right)  =0\,,\text{ \ }\dot{g}\left(  \tau\right)
=if\left(  \tau\right)  \,,\text{ \ }\dot{\varphi}\left(  \tau\right)
=-\frac{\Xi}{\sqrt{2}}\left[  ig\left(  \tau\right)  \sin^{2}\alpha+\tau
f\left(  \tau\right)  \cos^{2}\alpha\right]  \,, \label{2.6}%
\end{equation}
whose solutions has the form%
\begin{equation}
f\left(  \tau\right)  =f_{0}\,,\text{ \ }g\left(  \tau\right)  =g_{0}%
+if_{0}\tau\,,\text{ \ }\varphi\left(  \tau\right)  =-\left[  ig\left(
\tau\right)  \sin^{2}\alpha+\frac{f_{0}\tau}{2}\right]  \frac{\Xi\tau}%
{\sqrt{2}}\,, \label{2.7}%
\end{equation}
with constants $f_{0}$ and $g_{0}$. If $f\left(
\tau\right)  $, $g\left(  \tau\right)  $ and $\varphi\left(  \tau\right)  $
have the indicated form, then for $\hat{S}$ and $\hat
{A}\left(  \tau\right)  $ a basis of common eigenvectors can be found, for example%
\begin{align}
&  \hat{S}\left\vert \zeta,\tau\right\rangle =\lambda_{\zeta}\left(
\tau\right)  \left\vert \zeta,\tau\right\rangle \,,\label{2.8a}\\
&  \hat{A}\left(  \tau\right)  \left\vert \zeta,\tau\right\rangle
=\zeta\left\vert \zeta,\tau\right\rangle \,, \label{2.8b}%
\end{align}
where $\zeta\in\mathbb{C}$ and $\lambda_{\zeta}\left(  \tau\right)  $ is an arbitrary time-dependent function. If in addition to condition (\ref{2.5}) we require that the operator $\hat{A}\left(  \tau\right)  $ and its adjoint $\hat{A}^{\dagger}\left( \tau\right)  $ satisfy the usual commutation relation for creation and annihilation operators,%
\[
\left[  \hat{A}\left(  \tau\right)  ,\hat{A}^{\dagger}\left(  \tau\right)
\right]  =1\,,
\]
then the constants $f_{0}$ and $g_{0}$ satisfy the additional relation%
\begin{equation}
\operatorname{Re}\left[  f_{0}g^{\ast}\left(  \tau\right)  \right]
=\operatorname{Re}\left(  f_{0}g_{0}^{\ast}\right)  =1\,, \label{3.1}%
\end{equation}
which will be put to use in what follows. Thus, by virtue of relations (\ref{2.7}) and (\ref{3.1}), $\left\vert \zeta,\tau\right\rangle $ is an eigenstate of the annihilation operator $\hat{A}\left(  \tau\right)  $. In the $q$%
-representation, $\Phi_{\zeta}\left(  q,\tau\right)  =\left\langle
q|\zeta,\tau\right\rangle $, Eq. (\ref{2.8b}) takes the form%
\begin{equation}
\left[  \sqrt{2}\varphi\left(  \tau\right)  +f_{0}q+g\left(  \tau\right)
\partial_{q}\right]  \Phi_{\zeta}\left(  q,\tau\right)  =\sqrt{2}\zeta
\Phi_{\zeta}\left(  q,\tau\right)  \,. \label{3.1.a}%
\end{equation}
The general solution of Eq. (\ref{3.1.a}) is given by the formula%
\begin{equation}
\Phi_{\zeta}\left(  q,\tau\right)  =\exp\left[  -\frac{f_{0}}{g\left(
\tau\right)  }\frac{q^{2}}{2}-\frac{\sqrt{2}\left(  \varphi\left(
\tau\right)  -\zeta\right)  }{g\left(  \tau\right)  }q+i\phi\left(
\tau\right)  \right]  \,, \label{3.1.b}%
\end{equation}
where $\phi\left(  \tau\right)  $ is an arbitrary time-dependent function. Solution (\ref{3.1.b}) satisfies the Schr{\"o}dinger equation provided that is corresponds to the zero eigenvalue in Eq. (\ref{2.8a}). This means that $\phi\left(  \tau\right)  $ satisfies the condition%
\begin{align*}
&  \phi\left(  \tau\right)  =\frac{i}{2}\ln g\left(  \tau\right)  -\int
\frac{Q\left(  \tau\right)  }{2}d\tau-i\ln\mathcal{N}\,,\\
&  Q\left(  \tau\right)  =\left(  \frac{i\sqrt{2}\varphi\left(  \tau\right)
}{g\left(  \tau\right)  }+\Xi\tau\cos^{2}\alpha\right)  ^{2}=\left(
1-\frac{if_{0}}{g\left(  \tau\right)  }\frac{\tau}{2}\right)  ^{2}\Xi^{2}%
\tau^{2}\,,
\end{align*}
where $\mathcal{N}$ is a normalization constant. Finally, if we impose the condition of square integrability on $\Phi_{\zeta}\left(  q,\tau\right)  $, the the states represented by the formula (\ref{3.1.b}) take the final form%
\begin{align}
&  \Phi_{\zeta}\left(  q,\tau\right)  =\exp\left[  \sqrt{2}\frac{q+\sqrt
{2}\operatorname{Re}\left(  \varphi\left(  \tau\right)  g^{\ast}\left(
\tau\right)  \right)  }{g\left(  \tau\right)  }\zeta-\frac{g^{\ast}\left(
\tau\right)  }{g\left(  \tau\right)  }\frac{\zeta^{2}}{2}-\frac{\left\vert
\zeta\right\vert ^{2}}{2}\right]  \Phi_{0}\left(  q,\tau\right)
\,,\nonumber\\
&  \Phi_{0}\left(  q,\tau\right)  =\frac{1}{\pi^{1/4}\sqrt{g\left(
\tau\right)  }}\exp\left\{  -\frac{f_{0}}{g\left(  \tau\right)  }\frac{q^{2}%
}{2}-\sqrt{2}\frac{\varphi\left(  \tau\right)  }{g\left(  \tau\right)
}q-\left\vert g\left(  \tau\right)  \right\vert ^{2}\left[  \operatorname{Re}%
\left(  \frac{\varphi\left(  \tau\right)  }{g\left(  \tau\right)  }\right)
\right]  ^{2}+\int\frac{\operatorname{Re}Q\left(  \tau\right)  }{2i}%
d\tau\right\}  \,, \label{3.2}%
\end{align}
The two-parameter family of functions $\Phi_{\zeta}\left(  q,\tau\right)  $, which depend on the parameters $f_{0}$ and $g_{0}$, is called a family of nonstationary, generalized coherent states (GCSs). In addition to being solutions of the Schr{\"o}dinger equation (\ref{2.8a}),
satisfy the main properties of CSs, namely they are eigenstates of the annihilation operator, described by Eq. (\ref{2.8b}), and also form an overcomplete normalized basis \cite{Bag2014,Bag2015}. For completeness, we note that the GCSs can be written in the usual form as
\[
\left\vert \zeta,\tau\right\rangle =\exp\left(  -\frac{\left\vert
\zeta\right\vert ^{2}}{2}\right)\sum_{n=0}^{\infty}
\frac{\zeta^{n}}{\sqrt{n!}}\left\vert n,\tau\right\rangle \,,\ \ \left\vert
n,\tau\right\rangle =\frac{\left[  \hat{A}^{\dagger}\left(  \tau\right)
\right]  ^{n}}{\sqrt{n!}}\left\vert 0,\tau\right\rangle \,.
\]

\subsection{Uncertainty relations\label{Sec2.2}}

For some systems, the CSs possess the remarkable property that they can be constructed as solutions of the Heisenberg and Robertson–Schr{\"o}dinger uncertainty relations. The GCSs considered above possess the same properties
and, in particular, they minimize the Heisenberg uncertainty relation for $\tau=0$ for the appropriate choice of the constants $f_{0}$ and $g_{0}$. To prove these properties, it is convenient to express the operators $\hat{q}$, $\hat{p}$ in
terms of annihilation and creation operators $\hat{A}\left(  \tau\right)  $,
$\hat{A}^{\dagger}\left(  \tau\right)  $:%
\begin{equation}
\hat{q}=\frac{g^{\ast}\hat{A}+g\hat{A}^{\dagger}-2\operatorname{Re}\left(
g^{\ast}\left(  \tau\right)  \varphi\left(  \tau\right)  \right)  }{\sqrt{2}%
}\,,\text{ \ }\hat{p}=\frac{f_{0}^{\ast}\hat{A}-f_{0}\hat{A}^{\dagger
}-2i\operatorname{Im}\left(  f_{0}^{\ast}\varphi\left(  \tau\right)  \right)
}{i\sqrt{2}}\,.\label{5.1}%
\end{equation}
We also employ Eq. (\ref{2.8b}) to calculate the mean values%
\begin{align}
&  \bar{q}\left(  \tau\right)  \equiv\left\langle \zeta,\tau|\hat{q}%
|\zeta,\tau\right\rangle =\left\langle \hat{q}\right\rangle =\sqrt
{2}\operatorname{Re}\left[  g^{\ast}\left(  \zeta-\varphi\right)  \right]
=\bar{q}_{0}+\bar{p}_{0}\tau+\frac{\Xi}{2}\tau^{2}\,,\nonumber\\
&  \bar{p}\left(  \tau\right)  \equiv\left\langle \zeta,\tau|\hat{p}%
|\zeta,\tau\right\rangle =\left\langle \hat{p}\right\rangle =\sqrt
{2}\operatorname{Im}\left[  f_{0}^{\ast}\left(  \zeta-\varphi\right)  \right]
=\bar{p}_{0}+\Xi\tau\sin^{2}\alpha\,,\nonumber\\
&  \bar{q}_{0}=\bar{q}\left(  0\right)  =\sqrt{2}\operatorname{Re}\left(
g_{0}^{\ast}\zeta\right)  \,,\ \ \bar{p}_{0}=\bar{p}\left(  0\right)
=\sqrt{2}\operatorname{Im}\left(  f_{0}^{\ast}\zeta\right)  \,.\label{5.3}%
\end{align}
We use a horizontal bar, as usual, to denote the mean values. As a result, $\zeta$ can be written in the form%
\begin{equation}
\zeta=\frac{f_{0}\bar{q}\left(  \tau\right)  +ig\left(  \tau\right)  \bar
{p}\left(  \tau\right)  }{\sqrt{2}}+\varphi\left(  \tau\right)  =\frac
{f_{0}\bar{q}_{0}+ig_{0}\bar{p}_{0}}{\sqrt{2}}\,,\label{5.4}%
\end{equation}
which follows either from solving system (\ref{5.3}) for $\zeta$ or from
the mean value of $\hat{A}\left(  \tau\right)  $, as expected.

Calculating the mean values of the operators $\hat{q}^{2}$ and $\hat{p}^{2}$:%
\begin{equation}
\bar{q}^{2}\left(  \tau\right)  =2\left[  \operatorname{Re}\left(  g^{\ast
}\left(  \zeta-\varphi\right)  \right)  \right]  ^{2}+\frac{\left\vert
g\right\vert ^{2}}{2}\,,\text{ \ }\bar{p}^{2}\left(  \tau\right)  =2\left[
\operatorname{Im}\left(  f_{0}^{\ast}\left(  \zeta-\varphi\right)  \right)
\right]  ^{2}+\frac{\left\vert f_{0}\right\vert ^{2}}{2}\,,\label{5.5}%
\end{equation}
we find expressions for the standard deviation of the coordinate $\sigma_{q}\left(  \tau\right)  $ and
momentum $\sigma_{p}\left(  \tau\right)  $:%
\begin{align}
&  \sigma_{q}\left(  \tau\right)  =\sqrt{\left\langle \left(  \hat
{q}-\left\langle \hat{q}\right\rangle \right)  ^{2}\right\rangle }%
=\frac{\left\vert g\left(  \tau\right)  \right\vert }{\sqrt{2}}=\sqrt
{\sigma_{q}^{2}+\sqrt{4\sigma_{q}^{2}\sigma_{p}^{2}-1}\tau+\sigma_{p}^{2}%
\tau^{2}}\,,\nonumber\\
&  \sigma_{p}\left(  \tau\right)  =\sqrt{\left\langle \left(  \hat
{p}-\left\langle \hat{p}\right\rangle \right)  ^{2}\right\rangle }=\sigma
_{p}\left(  0\right)  \equiv\sigma_{p}=\frac{\left\vert f_{0}\right\vert
}{\sqrt{2}},\text{ \ }\sigma_{q}\equiv\sigma_{q}\left(  0\right)
=\frac{\left\vert g_{0}\right\vert }{\sqrt{2}}\,.\label{5.6}%
\end{align}
For the standard deviation $\sigma_{q}(\tau)$ in expression (\ref{5.6}) to be real, the Heisenberg uncertainty principle must be
fulfilled for $\tau=0$:%
\begin{equation}
\sigma_{q}\sigma_{p}\geq\frac{1}{2}\,.\label{5.7}%
\end{equation}
If this is true, then the uncertainty principle is fulfilled for any value of $\tau$ i. e., $\sigma_{q}\left(  \tau\right)  \sigma_{p}\left(  \tau\right)  \geq 1/2$, since $\sigma_{q}\left(  \tau\right)  \geq\sigma_{q}$. Moreover, calculating the covariance $\sigma_{qp}\left(  \tau\right)  $:%
\[
\sigma_{qp}\left(  \tau\right)  =\frac{1}{2}\left\langle \left(  \hat
{q}-\left\langle \hat{q}\right\rangle \right)  \left(  \hat{p}-\left\langle
\hat{p}\right\rangle \right)  -\left(  \hat{p}-\left\langle \hat
{p}\right\rangle \right)  \left(  \hat{q}-\left\langle \hat{q}\right\rangle
\right)  \right\rangle =\frac{f_{0}^{\ast}g\left(  \tau\right)  -1}{2i}\,,
\]
we find that the Robertson-Schr{\"o}dinger uncertainty relations \cite{Sch1930,Rob1930} are
 fulfilled identically identically:%
\begin{equation}
\sigma_{q}^{2}\left(  \tau\right)  \sigma_{p}^{2}\left(  \tau\right)
-\sigma_{qp}^{2}\left(  \tau\right)  =\frac{1}{4}\,.\label{5.18e}%
\end{equation}
This means that the GCSs are squeezed states.

In terms of the quantities $\bar{q}\left(  \tau\right)  $ and $\bar{p}\left(  \tau\right)  $ defined by Eqs. (\ref{5.3}), the GCS defined by Eqs. (\ref{3.2}) can be written in the form%
\begin{equation}
\Phi_{\zeta}\left(  q,\tau\right)  =\frac{1}{\pi^{1/4}\sqrt{g\left(
\tau\right)  }}\exp\left\{  -\frac{f_{0}}{g\left(  \tau\right)  }\frac{\left[
q-\bar{q}\left(  \tau\right)  \right]  ^{2}}{2}+\frac{i\bar{p}\left(
\tau\right)  }{2}\left[  2q-\bar{q}\left(  \tau\right)  \right]
+i\varrho\left(  \tau\right)  \right\}  \,, \label{5.16}%
\end{equation}
where%
\begin{align}
\varrho &  =\varrho^{\ast}=\frac{\operatorname{Im}\left(  f_{0}\varphi^{\ast
}\left(  \tau\right)  \right)  \bar{q}\left(  \tau\right)  +\operatorname{Re}%
\left(  g^{\ast}\left(  \tau\right)  \varphi\left(  \tau\right)  \right)
\bar{p}\left(  \tau\right)  }{\sqrt{2}}\nonumber\\
&  +\operatorname{Im}\left(  \frac{g^{\ast}\left(  \tau\right)  }{g\left(
\tau\right)  }\frac{\varphi^{2}\left(  \tau\right)  }{2}\right)  +\int
\frac{\operatorname{Re}Q\left(  \tau\right)  }{2}d\tau\,. \label{5.17}%
\end{align}
From this representation it is easy to see that the
corresponding probability density $\rho\left(  q,\tau\right)  $%
\begin{equation}
\rho\left(  q,\tau\right)  =\left\vert \Phi_{\zeta}\left(  q,\tau\right)
\right\vert ^{2}=\frac{1}{\sqrt{2\pi}\sigma_{q}\left(  \tau\right)  }%
\exp\left\{  -\frac{\left[  q-\bar{q}\left(  \tau\right)  \right]  ^{2}%
}{2\sigma_{q}^{2}\left(  \tau\right)  }\right\}  \,, \label{5.18}%
\end{equation}
is a Gaussian distribution, whose standard deviation $\sigma_{q}\left(
\tau\right)  $ corresponds to the standard deviation of the coordinates found in Eq. (\ref{5.6}). Therefore, the mean values $\bar
{q}\left(  \tau\right)  $, $\bar{p}\left(  \tau\right)  $ move along classical trajectories, which allows us to establish a one-to-one correspondence between the mathematical expectations given by Eqs. (\ref{5.6}) and the solutions of the Hamilton equations (\ref{1.5}). Employing Eq. (\ref{2.2}), one can write%
\begin{equation}
\bar{q}_{0}=\sqrt{2}\operatorname{Re}\left(  g_{0}^{\ast}\zeta\right)
=2\sigma_{q}\operatorname{Re}\left(  e^{-i\arg g_{0}}\zeta\right)
\,,\ \ \bar{p}_{0}=\sqrt{2}\operatorname{Im}\left(  f_{0}^{\ast}\zeta\right)
=2\sigma_{p}\operatorname{Im}\left(  e^{-i\arg f_{0}}\zeta\right)  \,,
\label{5.18b}%
\end{equation}
which allows us to find the correspondence%
\[
2\sigma_{z}\operatorname{Re}\left(  e^{-i\arg g_{0}}\zeta\right)
\leftrightarrow z_{0}\,,\ \ 2\sigma_{p_{z}}\operatorname{Im}\left(  e^{-i\arg
f_{0}}\zeta\right)  \leftrightarrow p_{z}^{0}\,,
\]
where $\sigma_{q}=l^{-1}\sigma_{z}$, $\sigma_{p}=l\hslash^{-1}\sigma
_{p_{z}}$.

If we impose the requirement of minimization of uncertainty relation (\ref{5.16}) for $\tau=0$
\begin{equation}
\sigma_{q}\sigma_{p}=\frac{1}{2}\,, \label{5.18c}%
\end{equation}
then the two-parameter family of GCSs described by Eq. (\ref{5.16}) and parameterized by $f_{0}$ and $g_{0}$, reduces to a oneparameter family with $f_{0}=f_{0}^{\ast}=g_{0}^{-1}$. The states obtained in this way are parameterized only by the initial standard deviations of the coordinates $\sigma_{q}$. For simplicity, we call these states coherent states (CSs).

\section{Coherent states as semiclassical ones\label{Sec3}}

CSs are Gaussian, and their density moves along a classical trajectory. These two characteristics are general for both coherent and semiclassical states. Therefore, it makes sense to ask whether coherent states can be assumed to be semiclassical at all times. In general, the answer is no since for semiclassical motion, variation in time of the corresponding probability density $\rho\left(  q,\tau\right)  $ should be slow. This means that quantum motion in some sense is bounded, and restricted to a region defined by the standard deviation of the coordinate $\sigma_{q}\left(  \tau\right)  $. In turn, this quantity depends on the momentum of the particle and on the external field; therefore, $\sigma_{q}\left(  \tau\right)  $ and $\rho\left(  q,\tau\right)  $can vary more rapidly or more slowly, depending on these quantities. In order to find the conditions under which the GCSs can be considered as SSs, let us turn to a criterion which is especially useful for one-dimensional motion\footnote{It should be noted that the given criterion is not general. One of the adopted definitions follows from the inequality $\zeta\gg1$, which is useful both for a particle moving in more than one dimension and for a particle in the so-called magnetic solenoid field \cite{Bag2014})}.

This criterion can be formulated as follows: the evolution of the density
$\rho\left(  q,\tau\right)  $ corresponds to the motion of the particle, but the standard deviation $\sigma_{q}\left(  \tau\right)  $ grows with time. Thus,
at some time $\tau$, the distribution for $\rho\left(  q,\tau\right)  $ has
``\textit{spread out}'' $\Delta\sigma
_{q}\left(  \tau\right)  =\sigma_{q}\left(  \tau\right)  -\sigma_{q}$ while the particle has passed the distance $\Delta z\left(  t\right)  =z\left(
t\right)  -z_{0}$. Writting $\Delta\sigma_{q}\left(  \tau\right)  $ in the
dimensional form $\Delta\sigma_{q}\left(  \tau\right)  =l^{-1}\Delta\sigma
_{z}\left(  t\right)  $:%
\begin{equation}
\Delta\sigma_{z}\left(  t\right)  =\sigma_{z}\left[  \sqrt{1+2\frac
{t}{t_{\sigma}}\sqrt{1-\left(  \frac{\hslash}{2\sigma_{z}\sigma_{p_{z}}%
}\right)  ^{2}}+\frac{t^{2}}{t_{\sigma}^{2}}}-1\right]  \,,\label{6.0}%
\end{equation}
where $t_{\sigma}=m\left(  \sigma_{z}/\sigma_{p_{z}}\right)  $, it is possible to compare the two length scales, using the ratio
\begin{equation}
R\left(  t\right)  =\frac{\Delta\sigma_{z}\left(  t\right)  }{\Delta z\left(
t\right)  }\,,\label{6.1}%
\end{equation}
and to discover that if $R\left(  t\right)  \ll1$, then the GCSs and the CSs can be
considered as SSs.

Let us examine the conditions under which the inequality $R\left(
t\right)  \leq1$ is fulfilled in the general case. We begin
with the case of a free particle. We set $\xi=0$
in relations (\ref{1.5}) and represent $R\left(  t\right)  $ in the form%
\begin{align}
&  R\left(  t\right)  =\frac{\sqrt{1+2\left(  t/t_{\sigma}\right)
\mathrm{Y}+\left(  t/t_{\sigma}\right)  ^{2}}-1}{\left(  t/t_{\sigma}\right)
\mathrm{X}}\,,\ \ \mathrm{X}=\frac{p_{z}}{\sigma_{p_{z}}}\,,\nonumber\\
&  \mathrm{Y}=\sqrt{1-\left(  \frac{\hslash}{2\sigma_{z}\sigma_{p_{z}}%
}\right)  ^{2}}=\sqrt{1-\frac{\mathrm{X}^{2}}{\mathrm{X}_{\sigma}^{2}}%
}\,,\ \ \mathrm{X}_{\sigma}=\frac{2p_{z}\sigma_{z}}{\hslash}\,.\label{6.2}%
\end{align}
From these relations it follows that $R\left(  t\right)  \leq1$ if one of the conditions%
\begin{equation}
\mathrm{i)\ X}\geq1\ \ \text{or \ }\mathrm{ii)}\ t<t_{\mathrm{c}}%
\mathrm{\ \ }\text{if\ }\mathrm{Y}\leq\mathrm{X}<1\,,\label{6.3}%
\end{equation}
is fulfilled, where $t_{\mathrm{c}}$ is the critical time for a free particle%
\begin{equation}
t_{\mathrm{c}}=\frac{\delta t_{\sigma}}{\left\vert \mathrm{X}^{2}-1\right\vert
}\,,\ \ \delta=2\left\vert \mathrm{X}-\mathrm{Y}\right\vert \,.\label{6.4}%
\end{equation}
Condition $(\mathrm{i)}$ means that the inequality $R\left(  t\right)  \leq1$ is fulfilled for any time $t$ as a consequence of the fact that the
momentum $p_{z}$ of the particle is greater than or equal to its standard deviation $\sigma_{p_{z}}$. In case \textrm{(ii)}, the inequality $R\left(  t\right)  \leq1$ is valid only during some intermediate time $t\in\left[  0,t_{\mathrm{c}}\right]  $. These properties become more transparent if we
introduce the de-Broglie wavelength $\lambda=2\pi
\hslash/p_{z}$. In the language of the de-Broglie wavelength, \textrm{X} and \textrm{Y} take the form%
\begin{equation}
\mathrm{X}=\left(  \frac{2\pi}{\lambda}\right)  \frac{\hslash}{\sigma_{p_{z}}%
}\,,\ \ \mathrm{X}_{\sigma}=\left(  \frac{2\pi}{\lambda}\right)  2\sigma
_{z}\,.\label{6.4b}%
\end{equation}
Consequently, it follows from condition $\mathrm{i)}$ that $\lambda\sigma_{p_{z}}\leq
2\pi\hslash$, and if the standard deviation $\sigma_{p_{z}}$ of the momentum is
significantly less than the inverse of the de-Broglie wavelength, $\sigma_{p_{z}}%
\ll2\pi\hslash/\lambda$, , then GCSs can be considered for any
time $t$. On the other hand, if the
condition $\mathrm{(ii)}$ is fulfilled, then the GCSs cannot be considered as SSs for arbitrary $t$ and are such only under the condition $t\ll
t_{\mathrm{c}}$. Therefore, GCSs can be considered as SSs if only one of the following conditions is fulfilled:%
\begin{equation}
\sigma_{p_{z}}\ll\frac{2\pi\hslash}{\lambda}\ \ \text{or}\ \ t\ll
t_{\mathrm{c}}\ \ \text{if}\ \ \frac{2\pi\hslash}{\lambda}<\sigma_{p_{z}}%
\leq\frac{2\pi\hslash}{\lambda}\left[  1+\left(  \frac{\lambda}{4\pi\sigma
_{z}}\right)  ^{2}\right]  ^{1/2}\,.\label{6.6}%
\end{equation}
For values of $\sigma_{p_{z}}$ from the last interval, the GCSs are quantum states for $t\geq t_{\mathrm{c}}$. What is more, the GCSs are quantum states, provided%
\begin{equation}
\sigma_{p_{z}}>\frac{2\pi\hslash}{\lambda}\left[  1+\left(  \frac{\lambda
}{4\pi\sigma_{z}}\right)  ^{2}\right]  ^{1/2}\,,\label{6.6b}%
\end{equation}
for any time $t\in\left[  0,\infty\right)  $.

The above arguments are somewhat different in the case of CSs. Taking equality
(\ref{5.18c}) into account, we have the equalities $\mathrm{X}=\mathrm{X}_{\sigma}$, $\mathrm{Y}=0$ , by virtue of which conditions (\ref{6.3}) are modified as follows:%
\begin{equation}
\mathrm{i)\ X}_{\sigma}\geq1\ \ \text{or}\ \ \mathrm{ii)}\ t<\tilde
{t}_{\mathrm{c}}\ \ \text{if}\ \ \mathrm{X}_{\sigma}<1\,, \label{6.7}%
\end{equation}
where the critical time $\tilde{t}_{\mathrm{c}}$ for a free particle is obtained as a particular case of Eq. (\ref{6.4}):%
\begin{equation}
\tilde{t}_{\mathrm{c}}=\frac{2t_{\sigma}\mathrm{X}_{\sigma}}{\left\vert
\mathrm{X}_{\sigma}^{2}-1\right\vert }\,. \label{6.8}%
\end{equation}
Consequently, CSs can be considered as SSs, provided the following conditions are satisfied:%
\begin{equation}
\lambda\ll4\pi\sigma_{z}\ \ \text{or}\ \ t\ll\tilde{t}_{\mathrm{c}%
}\ \ \text{if}\ \ \lambda\geq4\pi\sigma_{z}\,, \label{6.9}%
\end{equation}
which corresponds to a modification of the conditions (\ref{6.6}). It should be noted that the first of inequalities (\ref{6.9}) coincides with an earlier estimate \cite{Bag2014}. In the case $\lambda\geq4\pi\sigma_{z}$ the
CSs are quantum states for $t\geq\tilde{t}_{\mathrm{c}}$.

Let us consider the case of a particle in an electric field. According to the Hamilton equations (\ref{1.5}),%
\begin{equation}
\Delta z\left(  t\right)  =\sigma_{z}\left[  \left(  t/t_{\sigma}\right)
\mathrm{X}+\left(  t/t_{\sigma}\right)  ^{2}\mathrm{W}\right]
\,,\ \ \mathrm{W}=\frac{1}{2}\left(  \frac{mc}{\sigma_{p_{z}}}\right)
^{2}\frac{\sigma_{z}}{\lambdabar_{e}}\frac{E}{E_{\mathrm{c}}}\,, \label{6.10}%
\end{equation}
relation (\ref{6.1}) now has the form%
\begin{equation}
R\left(  t\right)  =\frac{\sqrt{1+2\left(  t/t_{\sigma}\right)  \mathrm{Y}%
+\left(  t/t_{\sigma}\right)  ^{2}}-1}{\left(  t/t_{\sigma}\right)
\mathrm{X}+\left(  t/t_{\sigma}\right)  ^{2}\mathrm{W}}\,, \label{6.11}%
\end{equation}
where $\lambdabar_{e}=\hslash/mc$ and $E_{\mathrm{c}}=m^{2}c^{3}/e\hslash$
are the Compton wavelength and the critical value of the electric field,
respectively. The electric field can be classified as strong if $2\mathrm{W}\geq1$, moderate for $1-\mathrm{X}^{2}\leq2\mathrm{W}<1$, and
weak ifn $0<2\mathrm{W}<1-\mathrm{X}^{2}$. Therefore, the inequality $R\left(
t\right)  \leq1$ is fulfilled for arbitrary $t\in\left[  0,\infty\right)
$ if just one of the following conditions holds:%
\begin{equation}
\mathrm{iii)\ X}\geq1\ \ \text{or}\ \ \mathrm{iv)\ }2\mathrm{W}\geq
1\ \ \text{or\ \ }\mathrm{v)\ Y}\leq\mathrm{X}<1\text{\ \ if\ \ }%
1-\mathrm{X}^{2}\leq2\mathrm{W}<1\,. \label{6.12}%
\end{equation}
Condition \textrm{(iii)} coincides with condition \textrm{(i)} in Eq.
(\ref{6.3}) for a free particle and does not depend on the amplitude $E$ of the electric field. Moreover, if the field is strong in accordance with condition $\mathrm{(iv)}$, then the inequality $R\left(  t\right)  \leq1$ is valid irrespective of the values of$\mathrm{X}$ and \textrm{Y } (or, equivalently, irrespective of the values of the standard deviation $\sigma_{p_{z}}$ of the momentum or standard deviation $\sigma_{z}$ of the coordinate). For fields with moderate amplitude, satisfying condition \textrm{(v)}, the inequality $R\left(  t\right)  \leq1$ is fulfilled only for $\mathrm{Y}\leq\mathrm{X}<1$. It should be noted that $R\left( t\right)  \leq1$ for arbitrary $t$ also for $\mathrm{X}=\mathrm{Y}$. This condition is impossible for a free particle, as is clear from Eqs. (\ref{6.2}), but can be fulfilled for a moderate electric field.

The cases when conditions (\ref{6.12}) are not fulfilled require separate analysis. For example, in the special case $\mathrm{X}=\mathrm{Y}$ and weak
electric fields%
\begin{equation}
\mathrm{vi)\ X}=\mathrm{Y}\text{ and }0<2\mathrm{W}<1-\mathrm{X}^{2}\,,
\label{6.13}%
\end{equation}
the inequality $R\left(  t\right)  \leq1$ is valid only for $t\geq t_{\mathrm{c}}^{\mathrm{vi}}:$%
\begin{equation}
t_{\mathrm{c}}^{\mathrm{vi}}=t_{\sigma}\frac{\mathrm{X}}{\mathrm{W}}\left(
\sqrt{1+\frac{\Delta}{\mathrm{X}^{2}}}-1\right)  \,,\ \ \Delta=\left\vert
\mathrm{X}^{2}+2\mathrm{W}-1\right\vert \,, \label{6.14}%
\end{equation}
where $t_{\mathrm{c}}^{\mathrm{vi}}$ is the critical time corresponding to
the condition \textrm{(iv)}. If conditions (\ref{6.12}) are not fulfilled, for example, in the cases%
\begin{equation}
\mathrm{vii)\ X}<\mathrm{Y\ \ }\text{and}\ \ 1-\mathrm{X}^{2}\leq
2\mathrm{W}<1\ \ \text{or}\ \ \mathrm{viii)\ X}<\mathrm{Y\ \ }\text{and}%
\ \ 0<2\mathrm{W}<1-\mathrm{X}^{2}\,, \label{6.15}%
\end{equation}
the critical time does not coincide with expression (\ref{6.14}) and, generally speaking, corresponds to a real positive solution of some cubic inequality. Under these conditions, it is more suitable to introduce a \textit{reference time} $t_{\mathrm{ref}}$ (different from the critical time), for which the condition $t\geq t_{\mathrm{ref}}$ is sufficient (even if not necessary) for the inequality $R\left(  t\right)  \leq1$ to be fulfilled. For example, under the condition \textrm{(vii)} the reference time is given by%
\begin{equation}
t_{\mathrm{ref}}^{\mathrm{vii}}=t_{\sigma}\min\left(  \frac{\delta}{\Delta
},\sqrt{\frac{\delta}{2\mathrm{XW}}},\left(  \frac{\delta}{\mathrm{W}^{2}%
}\right)  ^{1/3}\right)  \,,\ \ \delta=2\left\vert \mathrm{X}-\mathrm{Y}%
\right\vert \,, \label{6.16}%
\end{equation}
and in case \textrm{(viii)}%
\begin{equation}
t_{\mathrm{ref}}^{\mathrm{viii}}=t_{\sigma}\frac{\Delta}{4\mathrm{XW}}\left[
1+\sqrt{1+\frac{8\mathrm{XW}\delta}{\Delta^{2}}}\right]  \,. \label{6.17b}%
\end{equation}

The above results allow us to conclude that GCSs can be considered to be SSs for arbitrary $t$ if the standard deviation $\sigma_{p_{z}}$ of the momentum is much less than the inverse of the de-Broglie wavelength or if the electric field is sufficiently strong:%
\begin{equation}
\sigma_{p_{z}}\ll\frac{2\pi\hslash}{\lambda}\ \ \text{or\ \ }E\gg\left(
\frac{\sigma_{p_{z}}}{mc}\right)  ^{2}\left(  \frac{\lambdabar_{e}}{\sigma
_{z}}\right)  E_{\mathrm{c}}\,. \label{6.17c}%
\end{equation}
If neither of these conditions is fulfilled, then the GCSs cannot be considered as SSs for all $t$. For example, if the electric field is moderate in accordance with the conditions%
\begin{equation}
\mathrm{v)}\ \left(  \frac{\sigma_{p_{z}}}{mc}\right)  ^{2}\frac
{\lambdabar_{e}}{\sigma_{z}}\left[  1-\left(  \frac{2\pi\hslash}{\sigma
_{p_{z}}\lambda}\right)  ^{2}\right]  \leq\frac{E}{E_{\mathrm{c}}}<\left(
\frac{\sigma_{p_{z}}}{mc}\right)  ^{2}\frac{\lambdabar_{e}}{\sigma_{z}}\,,
\label{6.17c1}%
\end{equation}
and the corresponding standard deviations $\sigma_{p_{z}}$ and 
$\sigma_{z}$ satisfy the inequalities%
\begin{equation}
\mathrm{v)}\ \frac{2\pi\hslash}{\lambda}<\sigma_{p_{z}}\leq\frac{2\pi\hslash
}{\lambda}\left[  1+\left(  \frac{\lambda}{4\pi\sigma_{z}}\right)
^{2}\right]  ^{1/2}\,, \label{6.17c2}%
\end{equation}
then the GCSs can be considered as SSs only after the passage of a sufficiently long time, $t\gg0$. 

Stronger conditions arise for a weak field in the case when $\sigma_{p_{z}}$ and $\sigma_{z}$ satisfy the condition%
\begin{equation}
\mathrm{vi)}\ \sigma_{p_{z}}=\frac{2\pi\hslash}{\lambda}\left[  1+\left(
\frac{\lambda}{4\pi\sigma_{z}}\right)  ^{2}\right]  ^{1/2}\ \ \text{and}%
\ \ 0<\frac{E}{E_{\mathrm{c}}}<\left(  \frac{\sigma_{p_{z}}}{mc}\right)
^{2}\frac{\lambdabar_{e}}{\sigma_{z}}\left[  1-\left(  \frac{2\pi\hslash
}{\sigma_{p_{z}}\lambda}\right)  ^{2}\right]  \,. \label{6.17e}%
\end{equation}
In this case, the GCSs can be considered as SSs only for $t\gg t_{\mathrm{c}%
}^{\mathrm{vi}}$, where $t_{\mathrm{c}}^{\mathrm{vi}}$ is defined by Eq.
(\ref{6.14}). For sufficiently large values of $\sigma_{p_{z}}$%
\begin{equation}
\sigma_{p_{z}}>\frac{2\pi\hslash}{\lambda}\left[  1+\left(  \frac{\lambda
}{4\pi\sigma_{z}}\right)  ^{2}\right]  ^{1/2}\,, \label{6.17f}%
\end{equation}
and moderate electric fields, as in conditions (\ref{6.17c1}), the GCSs can be considered as SSs only for $t\gg t_{\mathrm{ref}}^{\mathrm{vii}}$, where
$t_{\mathrm{ref}}^{\mathrm{vii}}$ is assigned with the help of Eqs. (\ref{6.16}). However, if the electric field is weak, then the GCSs can be considered as SSs only for $t\gg t^{\mathrm{viii}}_{\mathrm{ref}}$, where $t^{\mathrm{viii}}_{\mathrm{ref}}$ is given by formula (\ref{6.17b}). These more rigid conditions are a consequence of inequalities (\ref{6.15}).

The above analysis must be modified if the considered states are CSs. In this case, $\mathrm{X}=\mathrm{X}_{\sigma}$ and $\mathrm{Y}=0$, where 
$\mathrm{W}=\mathrm{W}_{\sigma}=2\left(  \sigma_{z}/\lambdabar_{e}\right)
^{3}E/E_{\mathrm{c}}$, so that the conditions $\mathrm{(iii)}$, given by formula (\ref{6.12}), change over to%
\begin{equation}
\mathrm{iii)\ X}_{\sigma}\geq1\,,\ \ \mathrm{iv)\ }2\mathrm{W}_{\sigma}%
\geq1\,,\ \ \mathrm{v)\ X}_{\sigma}<1\ \ \text{and\ \ }1-\mathrm{X}_{\sigma
}^{2}\leq2\mathrm{W}_{\sigma}<1\,. \label{6.17}%
\end{equation}
Therefore, if just one of these conditions is satisfied, we will have $R\left(  t\right)  \leq1$ for any $t\in\left[  0,\infty\right)  $. If the
electric field is weak, but $\mathrm{X}_{\sigma}$ is a small quantity,%
\begin{equation}
\mathrm{ix)\ }0<2\mathrm{W}_{\sigma}<1-\mathrm{X}_{\sigma}^{2}\ \ \text{and}%
\ \ \mathrm{X}_{\sigma}<1\,, \label{6.18}%
\end{equation}
then the condition $R\left(  t\right)  \leq1$ will be valid for $t\geq
t_{\mathrm{ref}}^{\mathrm{ix}}$, where $t_{\mathrm{ref}}^{\mathrm{ix}}$ is the reference time corresponding to condition (\ref{6.18})%
\begin{equation}
t_{\mathrm{ref}}^{\mathrm{ix}}=t_{\sigma}\frac{\sqrt{\Delta_{\sigma}}%
}{\mathrm{W}_{\sigma}}\min\left(  1,\frac{\sqrt{\Delta_{\sigma}}}%
{2\mathrm{X}_{\sigma}}\right)  \,. \label{6.19}%
\end{equation}
Finally, in accordance with the above results, CSs can be considered as SSs if the standard deviation $\sigma
_{z}$ of the coordinates of the particle is much greater than its de-Broglie wavelenght $\lambda$ or if the
electric field is sufficiently strong:%
\begin{equation}
\mathrm{iii)\ }\sigma_{z}\gg\frac{\lambda}{4\pi}\ \ \text{or}%
\ \ \mathrm{iv)\ }E\gg\frac{1}{2}\left(  \frac{\lambdabar_{e}}{\sigma_{z}%
}\right)  ^{3}E_{\mathrm{c}}\,. \label{6.22}%
\end{equation}
If, however, the mean standard deviation of the coordinates is small, but the electric field is weak,%
\begin{equation}
\mathrm{ix)\ }0<\frac{E}{E_{\mathrm{c}}}<\frac{1}{4}\left(  \frac
{\lambdabar_{e}}{\sigma_{z}}\right)  ^{3}\left[  1-\left(  \frac{4\pi
\sigma_{z}}{\lambda}\right)  ^{2}\right]  \ \ \text{and}\ \ \sigma_{z}%
\leq\frac{\lambda}{4\pi}\,, \label{6.23}%
\end{equation}
then the CSs can, as before, be considered to be SSs for sufficiently large $t$ specifically for $t\gg t_{\mathrm{ref}}^{\mathrm{ix}}$.

\section{Conclusions}

In this work we have considered the problem of constructing GCSs and SSs for a nonrelativistic spinless particle in a constant uniform electric field. Use of the method of integrals of motion allowed us to analyze GCSs, the uncertainty relations, and the conditions under which these states can be considered as semiclassical. The obtained
GCSs satisfy the Robertson–Schr{\"o}dinger and Heisenberg uncertainty relations and in this their properties coincide with the GCSs of a free particle \cite{Bag2014}. Taking into account that the probability density for a GCS should vary slowly in time
for semiclassical motion, we have found conditions on the standard deviations of the coordinate $\sigma_{z}$ and the momentum $\sigma_{p_{z}}$and on the amplitude of the electric field, for which the GCSs are considered as SSs. Depending on the values of these quantities, CSs either can or cannot be considered as semiclassical for all time. In an examination of these conditions for a free particle, the GCSs can be considered as SSs, assuming that the de-Broglie wavelength of the electron satisfies the inequality $\lambda\ll 2\pi\hslash/\sigma_{p_{z}},$ while CSs can be considered as SSs if $\lambda\ll4\pi\sigma_{z}$. Both of these cases
are valid for all time. On the other hand, if
if $\lambda\geq2\pi\hslash/\sigma_{p_{z}}$ (for GCSs) or $\lambda\geq4\pi
\sigma_{z}$ (for CS), then (generalized) CSs can be considered as SSs only during a short interval of time, specifically for $t\ll t_{\mathrm{c}}$ (for the GCSs) or
$t\ll\tilde{t}_{\mathrm{c}}$ (for CSs). As for a particle interacting with an electric field, under these conditions the corresponding states can be considered as
SSs. However, fulfillment of these conditions is not required if the electric field is sufficiently strong (it satisfies the inequalities $E\gg\left(
\sigma_{p_{z}}/mc\right)  ^{2}\left(  \lambdabar_{e}/\sigma_{z}\right)
E_{\mathrm{c}}$ (for GCSs) or $2E\gg\left(  \lambdabar_{e}/\sigma
_{z}\right)  ^{3}E_{\mathrm{c}}$ (for CSs). For weak electric fields and
some values of the standard deviations, GCSs or CSs can be considered as SSs only after the passage of some interval of time. Thus, for a nonrelativistic spinless particle in a constant uniform electric field, GCSs or CSs can always be considered as SSs at sufficiently large times. At the same time, for a free particle this is possible only under some special conditions.

The authors express their gratitude to D. M. Gitman and V. G. Bagrov for helpful discussions.

This work was performed with financial support from the Program for Enhancement of Competitiveness of National Research Tomsk State University among the World’s Leading Research and Education Centers (T. C. Adorno) and from the Program for the Enhancement of Competitiveness of National Research Tomsk Polytechnic University among the World’s Leading Research and Education Centers (A. S. Pereira; Grant No. VIU-FTI-72/2017).

\end{document}